\documentclass[a4paper, 12pt, oneside]{article}
\usepackage[cp1251]{inputenc}
\usepackage[english, russian]{babel}
\usepackage{ graphics,amsfonts, amsmath,bm,amssymb, array}
\usepackage[dvips]{graphicx}
\renewcommand{\Re}{\mathop{\rm Re\,}}
\renewcommand{\Im}{\mathop{\rm Im\,}}

\makeatother {

\begin{document}
\thispagestyle{empty} \large
\renewcommand{\abstractname}{}
\renewcommand{\abstractname}{Abstract }
\renewcommand{\refname}{\begin{center} REFERENCES\end{center}}

 \begin{center}
\bf Transverse electric conductivity in quantum degenerate
collisional plasma in Mermin approach
\end{center}\medskip
\begin{center}
  \bf A. V. Latyshev\footnote{$avlatyshev@mail.ru$} and
  A. A. Yushkanov\footnote{$yushkanov@inbox.ru$}
\end{center}\medskip

\begin{center}
{\it Faculty of Physics and Mathematics,\\ Moscow State Regional
University, 105005,\\ Moscow, Radio str., 10--A}
\end{center}\medskip

\begin{abstract}
Formulas for transverse conductance in
quantum collisional plasma are deduced. The kinetic
equation in momentum space in the relaxation
approach is used. It is shown, that at $\hbar\to 0$ the derived formula
transfers to the classical one.
 It is shown also, that when electron collision frequency
tends to null (i.e. plasma becomes
collisionless), the conductance formula
transfers in the known formula inferred earlier by Lindhard.


\end{abstract}

\begin{center}
{\bf 1. Введение}
\end{center}
В хорошо известной работе Мермина \cite {Mermin} на основе анализа
неравновесной
матрицы плотности в $\tau $ -- приближении было получено выражение для
продольной диэлектрической проницаемости квантовой столкновитешльной плазмы.

Ранее в работе Линдхарда \cite{Lin} были получены выражения для
продольной и поперечной диэлектрической проницаемости квантовой
бесстоллкновительной плазмы. Затем Кливер и Фукс показали
\cite{Kliewer}, что прямое обобщение формул Линдхарда на случай
столкновительной плазмы (путем замены $\omega\to \omega+i/\tau$)
некорректно. Этот недостаток для продольной диэлектрической
проницаемости был преодолен в работе Мермина \cite{Mermin}. В тоже
самое время до настоящего времени не имеется корректного выражения
для поперечной диэлектрической проницаемости для случая вырожденной
квантовой столкновительной плазмы. Цель настоящей работы
--- восполнить этот пробел.

Свойства электрической проводимости и диэлектрической проницаемости по
формулам, выведенным Линдхардом \cite{Lin}, подробно изучались в монографии
\cite{Dressel}. В работе \cite{Gelder} поперечная
диэлектрическая проницаемость
квантовой плазмы применялась в вопросах теори скин--эффекта.
В настоящее время растет интерес к изучению различных свойств квантовой
плазмы (см, например, \cite{Anderson}--\cite{Manf2}). Особенно следует
отметить работу Дж. Манфреди \cite{Manf}, посвященную исследованию
электромагнитных свойств квантовой плазмы.

Диэлектрическая проницаемость плазмы определяется электрической
проводимостью плазмы. Поэтому сначала мы рассмотрим поперечную электрическую
проводимость квантовой столкновительной плазмы.

\begin{center}
  \bf 1. Кинетическое уравнение для матрицы плотности
\end{center}

Пусть векторный потенциал электромагнитного поля является гармоническим, т.е.
изменяется как
$
{\bf A}={\bf A}({\bf r})\exp(-i \omega t).
$
Мы рассматриваем поперечную проводимость. Поэтому выполняется следующее
соотношение
$
{\bf \rm \bf div A}(\mathbf{r},t)=0.
$
Связь между векторным потенциалом и напряженностью электрического поля
дается следующим выражением
$$
{\bf A}({\bf q})=-\dfrac{ic}{\omega}\;{\bf E}({\bf q}).
$$

Равновесная матрица плотности имеет следующий вид
$$
{\tilde \rho}=\Theta(E_F-H).
$$

Здесь $\Theta(x)$ -- функция Хэвисайда,
$$
\Theta(x)=\left\{\begin{array}{c}
            1,\qquad x>0, \\
            0,\qquad x<0,
          \end{array}\right.
$$
$E_F$ -- энергия электрона на поверхности Ферми, которая
считается сферической, $E_F=\dfrac{mv_F^2}{2}$, $v_F$ -- скорость
электрона на поверхности Ферми, $m$ -- масса электрона,
$H$ -- гамильтониан.

В линейном приближении гамильтониан имеет следующий вид
$$
H=\dfrac{({\bf p}-({e}/{c}){\bf A})^2}{2m}=
\dfrac{{\bf p}^2}{2m}-\dfrac{e}{2mc}({\bf p}{\bf A}+{\bf A} {\bf
p}).
$$

Здесь $\mathbf{p}$ -- оператор импульса, $\mathbf{p}=-i\hbar \nabla$,
$e$ -- заряд электрона, $c$ -- скорость света.

Следовательно, мы можем представить этот гамильтониан в виде суммы
двух операторов $H=H_0+H_1$, где
$$
H_0=\dfrac{{\bf p}^2}{2m},
\qquad H_1=-\dfrac{e}{2mc}({\bf p}{\bf A}+{\bf A} {\bf p}).
$$

Возьмем кинетическое уравнение фон Неймана---Больцмана для
матрицы плотности
$$
i\hbar\dfrac{\partial \rho}{\partial t}=[H,\rho]+i\hbar
B[\rho,\rho],
$$
где $B[\rho,\rho]$ -- нелинейный интеграл столкновений
Больцмана, $[H,\rho]=H\rho-\rho H$ -- коммутатор, $\hbar$ --
постоянная Больцмана.

Заменим в этом уравнении нелинейный интеграл столкновений $B[\rho,\rho]$ на
модельный интеграл столкновений релаксационного типа
$$
M[\rho]=\dfrac{\tilde{\rho}-\rho}{\tau}.
$$

Здесь $\nu=1/\tau$ -- эффективная частота столкновений частиц плазмы,
$\tau$ -- характерное время между двумя последовательными столкновениями,
$\mathbf{\tilde{\rho}}$ -- равновесная матрица плотности.

Такой интеграл столкновений называют интегралом столкновений
типа БГК (Бхатнагар, Гросс, Крук) \cite{BGK}. Независимо от \cite{BGK}
этот интеграл столкновений был введен в \cite{Welander}.

В результате указанной замены получаем эволюционное кинетическое
уравнение, введенное Мерминым \cite{Mermin}
Возьмем кинетическое уравнение для матрицы плотности в $\tau$ -- приближении
$$
i\hbar \dfrac{\partial \rho}{\partial t}=[H,\rho]+
{i\hbar}\dfrac{{\tilde\rho}-\rho}{\tau}.
\eqno{(1.1)}
$$

В линейном приближении по внешнему полю мы ищем матрицу плотности в виде
$$
{\rho}={\tilde \rho}^{(0)}+{\rho}^{(1)}.
\eqno{(1.2)}
$$

Здесь ${\rho}^{(1)}$ -- поправка (возмущение) к равновесной
матрице плотности, обусловленная наличием электромагнитного поля,
$\tilde{\rho}^{(0)}$ -- равновесная матрица плотности,
отвечающая "равновесному"\, оператору Гамильтона $H_0$, т.е.
$\tilde{\rho}^{(0)}=\Theta(E_F-H_0)$.

Представим равновесную матрицу плотности $\tilde{\rho}$
в следующем виде
$$
\tilde{\rho}=\tilde{\rho}^{(0)}+\tilde{\rho}^{(1)}.
\eqno{(1.3)}
$$

Рассмотрим коммутатор $[H, \tilde{\rho}]$. В линейном приближении
этот коммутатор равен
$$
[H, {\tilde \rho}\,]=[H_0, {\tilde \rho}^{(1)}]+[H_1, {\tilde
\rho}^{(0)}]
\eqno{(1.4)}
$$
и
$$
[H, {\tilde \rho}\,]=0.
\eqno{(1.5)}
$$

Для коммутаторов из правой части равенства (1.4) мы находим
$$
\langle\mathbf{k}_1|[H_0, \tilde{\rho}^{(1)}]|\mathbf{k}_2\rangle=
\big(E_{\mathbf{k}_1}-E_{\mathbf{k}_2}\big)\tilde{\rho}^{(1)}
(\mathbf{k}_1-\mathbf{k}_2),
\eqno{(1.6)}
$$
и
$$
\langle\mathbf{k}_1|[H_1, \tilde{\rho}^{(0)}]|\mathbf{k}_2\rangle=
(f_{\mathbf{k}_2}-f_{\mathbf{k}_1})\langle\mathbf{k}_1|H_1|
\mathbf{k}_2\rangle=
$$
$$
=\dfrac{e}{2mc}(f_{\mathbf{k}_2}-f_{\mathbf{k}_1})
(\mathbf{k}_1+\mathbf{k}_2)\mathbf{A}(\mathbf{k}_1-\mathbf{k}_2),
\eqno{(1.7)}
$$
где
$$
f_{\mathbf{k}}=\Theta(E_F-E_{\mathbf{k}}),\quad
E_{\mathbf{k}}=\dfrac{\hbar^2\mathbf{k}^2}{2m}, \qquad
\mathbf{p}=\hbar \mathbf{k}.
$$

Из соотношений (1.3)--(1.7) вытекает, что
$$
\tilde{\rho}^{(1)}(\mathbf{k}_1-\mathbf{k}_2)=
-\dfrac{e\hbar}{2mc}\dfrac{f_{\mathbf{k}_1}-
f_{\mathbf{k}_2}}{E_{\mathbf{k}_1}-E_{\mathbf{k}_2}}
(\mathbf{k}_1+\mathbf{k}_2)\mathbf{A}(\mathbf{k}_1-\mathbf{k}_2).
\eqno{(1.8)}
$$

С помощью соотношений (1.2)--(1.4) мы линеаризуем кинетическое уравнение
(1.1).
Получаем следующее уравнение
$$
i\hbar \dfrac{\partial \rho^{(1)}}{\partial t}=[H_0,
\rho^{(1)}]+[H_1, \tilde{\rho}^{(0)}]+i\hbar \nu(\tilde{\rho}^{(1)}-
\rho^{(1)}).
\eqno{(1.9)}
$$

Заметим, что возмущение $\rho^{(1)}\sim \exp(-i\omega t)$,
Тогда уравнение (1.9) принимает следующий вид
$$
\hbar (\omega+i\nu)\rho^{(1)}=[H_0, \rho^{(1)}]+
[H_1, \tilde{\rho}^{(0)}]+i\hbar \nu \tilde{\rho}^{(1)}.
$$
Отсюда следует, что
$$
\hbar (\omega+i\nu)\langle\mathbf{k}_1|\rho^{(1)}|\mathbf{k}_2\rangle=
\langle\mathbf{k}_1|[H_0,\rho^{(1)}]|\mathbf{k}_2\rangle+
$$
$$
+\langle\mathbf{k}_1|[H_1,\tilde{\rho}^{(0)}]|\mathbf{k}_2\rangle+
i\hbar \nu\langle\mathbf{k}_1|\tilde{\rho}^{(1)}|\mathbf{k}_2\rangle.
\eqno{(1.10)}
$$

Нетрудно видеть, что
$$
\langle\mathbf{k}_1|[H_0,\rho^{(1)}]|\mathbf{k}_2\rangle=(E_{\mathbf{k}_1}-
E_{\mathbf{k}_2})\langle\mathbf{k}_1|\rho^{(1)}|\mathbf{k}_2\rangle=
$$
$$
=(E_{\mathbf{k}_1}-E_{\mathbf{k}_2})\rho^{(1)}(\mathbf{k}_1-\mathbf{k}_2).
\eqno{(1.11)}
$$

Подставим в (1.10) равенства (1.11) и (1.7). В результате будем
иметь:
$$
[\hbar(\omega+i\nu)-E_{\mathbf{k}_1}+E_{\mathbf{k}_2}]
\rho^{(1)}(\mathbf{k}_1-\mathbf{k}_2)=
$$
$$
=\dfrac{e\hbar}{2mc}(f_{\mathbf{k}_1}-f_{\mathbf{k}_2})
(\mathbf{k}_1+\mathbf{k}_2)\mathbf{A}(\mathbf{k}_1-\mathbf{k}_2)+
i\hbar \nu \tilde{\rho}^{(1)}(\mathbf{k}_1-\mathbf{k}_2).
$$

Преобразуем теперь уравнение (1.12) с использованием равенства
(1.8) к уравнению, из которого находим
$$
\rho^{(1)}(\mathbf{k}_1-\mathbf{k}_2)=
-\dfrac{e\hbar}{2mc}K(\mathbf{k}_1,\mathbf{k}_2)
(f_{\mathbf{k}_1}-f_{\mathbf{k}_2})
(\mathbf{k}_1+\mathbf{k}_2)\mathbf{A}(\mathbf{k}_1-\mathbf{k}_2).
\eqno{(1.12)}
$$
где
$$
K(\mathbf{k}_1,\mathbf{k}_2)=\dfrac{E_{\mathbf{k}_1}-
E_{\mathbf{k}_2}-i\hbar \nu}{(E_{\mathbf{k}_1}-E_{\mathbf{k}_2})
[E_{\mathbf{k}_1}-E_{\mathbf{k}_2}-\hbar(\omega+i \nu)]}.
$$
В уравнении (1.12) мы положим $\mathbf{k}_1=\mathbf{k}$,
$\mathbf{k}_2=\mathbf{k}-\mathbf{q}$. Тогда
$$
\langle\mathbf{k}_1|\rho^{(1)}|\mathbf{k}_2\rangle=
\langle\mathbf{k}|\rho^{(1)}|\mathbf{k}-
\mathbf{q}\rangle=\rho^{(1)}(\mathbf{q})=
$$ \hspace{0.2cm}
$$
=-\dfrac{e\hbar}{2mc}K(\mathbf{k},\mathbf{q})(f_{\mathbf{k}}-
f_{\mathbf{k-q}})\mathbf{k}\mathbf{A}(\mathbf{q}).
\eqno{(1.13)}
$$\hspace{0.3cm}

В (1.13) обозначено
$$
K(\mathbf{k},\mathbf{q})=\dfrac{E_{\mathbf{k}}-
E_{\mathbf{k-q}}-i\hbar \nu}{(E_{\mathbf{k}}-E_{\mathbf{k-q}})
[E_{\mathbf{k}}-E_{\mathbf{k-q}}-\hbar(\omega+i \nu)]}.
$$

\begin{center}
  \bf 2. Плотность тока
\end{center}

Плотность тока ${\bf j}=\mathbf{j}({\bf q,\omega,\nu})$ определяется как
$$
{\bf j}=e\int \dfrac{d{\bf k}}{8\pi^3m}\left\langle{\bf k}+
\frac{\mathbf{q}}{2}\left|({\bf p}-\frac{e}{c}{\bf A})
\rho+\rho ({\bf p}-\frac{e}{c}{\bf A} \big)\right|{\bf k}-
\frac{\mathbf{q}}{2}\right\rangle.
\eqno{(2.1)}
$$
Подставляя (1.3) в подынтегральное выражение из (2.1), имеем
$$
\left\langle{\bf k}+
\frac{\mathbf{q}}{2}\left|({\bf p}-\frac{e}{c}{\bf A})
\rho+\rho ({\bf p}-\frac{e}{c}{\bf A} \big)\right|{\bf k}-
\frac{\mathbf{q}}{2}\right\rangle=$$$$=
\left\langle{\bf k}+
\frac{\mathbf{q}}{2}\left|{\bf p}\rho^{(1)}+\rho^{(1)}\mathbf{p}-\frac{e}{c}
({\bf A}\tilde{\rho}^{(0)}+\tilde{\rho}^{(0)}\mathbf{A})\right|{\bf k}-
\frac{\mathbf{q}}{2}\right\rangle.
$$
Нетрудно показать, что
$$
\left\langle{\bf k}+\dfrac{\mathbf{q}}{2}\left|{\bf p}\rho^{(1)}+
\rho^{(1)}\mathbf{p}\right|{\bf k}-
\dfrac{\mathbf{q}}{2}\right\rangle
=2\hbar\mathbf{k}\rho^{(1)}(\mathbf{q}),
$$
и
$$
\left\langle{\bf k}+
\dfrac{\mathbf{q}}{2}\left|{\bf A}\tilde{\rho}^{(0)}+
\tilde{\rho}^{(0)}\mathbf{A}\right|{\bf k}-\dfrac{\mathbf{q}}{2}\right\rangle=
\mathbf{A}(\mathbf{q})\Big[\tilde{\rho}^{(0)}(\mathbf{k}+
\dfrac{\mathbf{q}}{2})+\tilde{\rho}^{(0)}(\mathbf{k}-
\dfrac{\mathbf{q}}{2})\Big].
$$

Следовательно, выражение для плотности тока имеет следующий вид
$$
\mathbf{j}=-\dfrac{e^2}{mc}\mathbf{A}(\mathbf{q})
\int \dfrac{d\mathbf{k}}{8\pi^3}\tilde{\rho}^{(0)}(\mathbf{k}+
\dfrac{\mathbf{q}}{2})-\dfrac{e^2}{mc}\mathbf{A}(\mathbf{q})
\int \dfrac{d\mathbf{k}}{8\pi^3}\tilde{\rho}^{(0)}(\mathbf{k}-
\dfrac{\mathbf{q}}{2})+
$$
$$
+e\hbar\int\dfrac{d\mathbf{k}}{4\pi^3m}
\left\langle\mathbf{k}+\dfrac{\mathbf{q}}{2}
\left|\rho^{(1)}\right|\mathbf{k}-\dfrac{\mathbf{q}}{2}\right\rangle.
$$

Первые два члена в этом выражении равны друг другу
$$
\int \dfrac{d\mathbf{k}}{8\pi^3}\tilde{\rho}^{(0)}(\mathbf{k}+
\dfrac{\mathbf{q}}{2})=
\int \dfrac{d\mathbf{k}}{8\pi^3}\tilde{\rho}^{(0)}(\mathbf{k}-
\dfrac{\mathbf{q}}{2})=\dfrac{N}{2},
$$
где $N$ -- числовая плотность (концентрация) плазмы.

Следовательно, плотность тока равна
$$
{\bf j}=-\frac{e^2N}{mc}{\bf A}({\bf q})+
e\hbar\int \dfrac{d{\bf k}}{4\pi^3m}{\bf k}
\left\langle\mathbf{k}+\dfrac{\mathbf{q}}{2}\left|\rho^{(1)}\right|{\bf k}-
\frac{\mathbf{q}}{2}\right\rangle.
\eqno{(2.2)}
$$

Первое слагаемое в (2.2) есть не что иное, как калибровочная плотность тока.

С помощью очевидной замены переменных  выражение (2.2)
можно преобразовать к виду
$$
{\bf j}=-\frac{e^2N}{mc}{\bf A}({\bf q})+
e\hbar\int \dfrac{d{\bf k}}{4\pi^3m}{\bf k}
\left\langle\mathbf{k}\left|\rho^{(1)}\right|{\bf k}-{\bf q}\right\rangle.
\eqno{(2.3)}
$$

В соотношении (2.3) подынтегральное выражение дается равенством (1.13).
Подставляя (1.13) в (2.3), получаем следующее выражение для плотности тока
$$
{\bf j}=-\frac{e^2N}{mc}{\bf A}({\bf q})
-\dfrac{e^2\hbar^2}{m^2c}\int \dfrac{\mathbf{k}d\mathbf{k}}{4\pi^3}
[\mathbf{k}\mathbf{A(q)}]
K(\mathbf{k},\mathbf{q})(f_{\mathbf{k}}-f_{\mathbf{k-q}}).
\eqno{(2.4)}
$$

Направим ось $x$ вдоль вектора ${\bf q}$, а ось $y$ вдоль
вектора ${\bf A}$. Тогда предыдущее векторное выражение (2.4) может быть
переписано в виде трех скалярных
$$
{ j}_y({\bf q},\omega,\nu)=-\frac{e^2N}{mc}{ A}({\bf q})-
\dfrac{e^2\hbar^2 A({\bf q})}{m^2c}\int \dfrac{d{\bf k}}{4\pi^3}{ k}_y^2
\;K(\mathbf{k},\mathbf{q})(f_{\mathbf{k}}-f_{\mathbf{k-q}})
$$
и
$$
{ j}_x({\bf q},\omega,\nu)={ j}_z({\bf q},\omega,\nu)=0.
$$

Очевидно, что
$$
\int \dfrac{d{\bf k}}{4\pi^3}{ k}_y^2\;
K(\mathbf{k},\mathbf{q})(f_{\mathbf{k}}-f_{\mathbf{k-q}})
=\int \dfrac{d{\bf k}}{4\pi^3}{ k}_z^2\;
K(\mathbf{k},\mathbf{q})(f_{\mathbf{k}}-f_{\mathbf{k-q}}).
$$
Следовательно
$$
\int \dfrac{d{\bf k}}{4\pi^3}{ k}_y^2\;
K(\mathbf{k},\mathbf{q})(f_{\mathbf{k}}-f_{\mathbf{k-q}})=
$$
$$
=\dfrac{1}{2}\int \dfrac{d{\bf k}}{4\pi^3}({ k}_y^2+{
k}_z^2)\;K(\mathbf{k},\mathbf{q})(f_{\mathbf{k}}-f_{\mathbf{k-q}})
\dfrac{1}{2}\int \dfrac{d{\bf k}}{4\pi^3}({\bf k}^2-{
k}_x^2)\;K(\mathbf{k},\mathbf{q})(f_{\mathbf{k}}-f_{\mathbf{k-q}}).
$$

Отсюда мы заключаем, что выражение для плотности тока можно представить в
следующей инвариантной форме
$$
{\bf j}=-\frac{e^2N}{mc}{\bf A}({\bf q})-
-\dfrac{e^2\hbar^2}{8\pi^3m^2c}{\bf A}({\bf q})\int d{\bf k}
K(\mathbf{k},\mathbf{q})(f_{\mathbf{k}}-f_{\mathbf{k-q}})\mathbf{k}_\perp^2.
\eqno{(2.5)}
$$
где
$$
\mathbf{k}_\perp^2={\bf k}^2-\Big(\dfrac{{\bf k}{\bf q}}{q}\Big)^2
$$

Представим ядро $K(\mathbf{k},\mathbf{q})$ в виде суммы двух
слагаемых
$$
K(\mathbf{k},\mathbf{q})=\dfrac{1}{E_{\mathbf{k}}-E_{\mathbf{k-q}}}+
\dfrac{\hbar \omega}{(E_{\mathbf{k}}-E_{\mathbf{k-q}})[E_{\mathbf{k}}-
E_{\mathbf{k-q}}-\hbar(\omega+i \nu)]}.
$$
С помощью этого разложения представим плотность тока в следующем
виде:
$$
\mathbf{j}=-\dfrac{e^2N}{mc}\mathbf{A}(\mathbf{q})-\dfrac{e^2\hbar^2}
{8\pi^3 mc}
\int \dfrac{f_{\mathbf{k}}-f_{\mathbf{k-q}}}{E_{\mathbf{k}}-
E_{\mathbf{k-q}}}\mathbf{k}_\perp^2d\mathbf{k}-
$$
$$
-\dfrac{e^2\hbar^3\omega}{8\pi^3m^2c}\mathbf{A}(\mathbf{q})\int
\dfrac{(f_{\mathbf{k}}-f_{\mathbf{k-q}})\mathbf{k}_\perp^2d\mathbf{k}}
{(E_{\mathbf{k}}-E_{\mathbf{k-q}})[E_{\mathbf{k}}-
E_{\mathbf{k-q}}-\hbar(\omega+i \nu)]}
\eqno{(2.6)}
$$

Первые два члена в предыдущем соотношении (2.6) не зависят от частоты
$\omega$ и определяются диссипативными свойствами материала, определяемыми
частотой столкновений $\nu$. Эти члены являются универсальными параметрами,
определяющими диамагнетизм Ландау.

\begin{center}
\bf 3. Поперечная электрическая проводимость и диэлектрическая проницаемость
\end{center}

Учитывая связь векторного потенциала с напряженностью электромагнитного поля,
а также связь плотности тока с электрическим полем, на основании предыдущего
соотношения (2.5) получаем следующее выражение инвариантного вида для поперечной
электрической проводимости
$$
\sigma_{tr}
=\dfrac{ie^2N}{m\omega}+\dfrac{ie^2\hbar^2}{8\pi^3m^2\omega}
\int K(\mathbf{k},\mathbf{q})(f_{\mathbf{k}}-f_{\mathbf{k-q}})
\mathbf{k}_\perp^2 d\mathbf{k}.
\eqno{(3.1)}
$$

Выделяя в (3.1) статическую проводимость $\sigma_0=e^2N/m \nu$,
перепишем (3.1) в виде
$$
\dfrac{\sigma_{tr}}{\sigma_0}=\dfrac{i\nu}{\omega}\Big[1+
\dfrac{\hbar^2}{8\pi^3mN}\int K(\mathbf{k},\mathbf{q})
(f_{\mathbf{k}}-f_{\mathbf{k-q}})\mathbf{k}_\perp^2
d\mathbf{k}\Big].
\eqno{(3.2)}
$$

На основании (3.2) напишем выражение для поперечной
диэлектрической проницаемости
$$
\varepsilon_{tr}=1-\dfrac{\omega_p^2}{\omega^2}\Big[1+
\dfrac{\hbar^2}{8\pi^3mN}\int K(\mathbf{k},\mathbf{q})
(f_{\mathbf{k}}-f_{\mathbf{k-q}})\mathbf{k}_\perp^2
d\mathbf{k}\Big].
\eqno{(3.3)}
$$

Если воспользоваться разложением ядра на элементарные дроби
$$
K(\mathbf{k},\mathbf{q})=\dfrac{i \nu}{\omega+i \nu}\dfrac{1}
{E_{\mathbf{k}}-E_{\mathbf{k-q}}}+\dfrac{\omega}{\omega+i \nu}\dfrac{1}
{E_{\mathbf{k}}-E_{\mathbf{k-q}}-\hbar (\omega+i\nu)},
$$
то для поперечной электрической проводимости (3.1) и диэлектрической
проницаемости (3.2) мы имеем следующие два явных
представления:
$$
\dfrac{\sigma_{tr}}{\sigma_0}=\dfrac{i \nu}{\omega}\Big(1+
\dfrac{\omega J_\omega+i \nu J_\nu}{\omega+i \nu}\Big)
\eqno{(3.4)}
$$\medskip
и
$$
\varepsilon_{tr}=1- \dfrac{\omega_p^2}{\omega^2}\Big(1+
\dfrac{\omega J_\omega+i \nu J_\nu}{\omega+i \nu}\Big).
\eqno{(3.5)}
$$\medskip

В равенствах (3.3) и (3.4) введены обозначения:
$$
J_\omega=\dfrac{\hbar^2}{8\pi^3mN}\int
\dfrac{(f_{\mathbf{k}}-f_{\mathbf{k-q}})\mathbf{k}_\perp^2
d\mathbf{k}}{E_{\mathbf{k}}-E_{\mathbf{k-q}}-\hbar (\omega+i\nu)}
$$
и
$$
J_\nu=\dfrac{\hbar^2}{8\pi^3mN}\int
\dfrac{f_{\mathbf{k}}-f_{\mathbf{k-q}}}{E_{\mathbf{k}}-E_{\mathbf{k-q}}}
\mathbf{k}_\perp^2 d\mathbf{k}.
$$

Преобразуем эти интегралы к следующему виду:
$$
J_\omega=\dfrac{\hbar^2}{8\pi^3mN}\int \dfrac{[2E_{\mathbf{k}}-
(E_{\mathbf{k-q}}-E_{\mathbf{k+q}})]f_{\mathbf{k}}\mathbf{k}_\perp^2
d\mathbf{k}}{[E_{\mathbf{k}}-E_{\mathbf{k-q}}-\hbar(\omega+i \nu)]
[E_{\mathbf{k}}-E_{\mathbf{k+q}}+\hbar(\omega+i \nu)]},
\eqno{(3.6)}
$$
$$
J_\nu=\dfrac{\hbar^2}{8\pi^3mN}\int \dfrac{2E_{\mathbf{k}}-
(E_{\mathbf{k-q}}-E_{\mathbf{k+q}})}{(E_{\mathbf{k}}-E_{\mathbf{k-q}})
(E_{\mathbf{k}}-E_{\mathbf{k+q}})}f_{\mathbf{k}}\mathbf{k}_\perp^2
d\mathbf{k}.
\eqno{(3.7)}
$$

Вместо вектора $\mathbf{k}$ введем безразмерный вектор $\mathbf{P}$
следующим равенством
$\mathbf{P}=\dfrac{\mathbf{k}}{k_F}$,
где $k_F$ -- волновое число Ферми, $k_F=\dfrac{mv_F}{\hbar}$.

Тогда
$$
\mathbf{k}_\perp^2d\mathbf{k}=(k^2-k_x^2)d^3k=k_F^5(P^2-P_x^2)d^3P=
k_F^5P_\perp^2d^3K,
$$
где
$$
P_\perp^2=P^2-P_x^2=P_y^2+P_z^2.
$$

Преобразуем интегралы (3.6) и (3.7). В этих интегралах
$$
E_{\mathbf{k}}=\dfrac{\hbar^2\mathbf{k}^2}{2m}=\dfrac{\hbar^2
k_F^2}{2m}P^2=\dfrac{mv_F^2}{2}P^2=E_FP^2,
$$
$$
E_{\mathbf{k\mp q}}=\dfrac{\hbar^2(\mathbf{k-q})^2}{2m}=
\dfrac{\hbar^2k_F^2}{2m}(\mathbf{P}\mp \mathbf{l})^2, \qquad
l=\dfrac{q}{k_F},
$$
$l$ -- безразмерное волновое число.

Теперь получаем
$$
E_{\mathbf{k}}-E_{\mathbf{k-q}}=2E_Fl(P_x-\dfrac{l}{2}),\quad
E_{\mathbf{k}}-E_{\mathbf{k+q}}=-2E_Fl(P_x+\dfrac{l}{2}),
$$

$$
2E_{\mathbf{k}}-(E_{\mathbf{k-q}}-E_{\mathbf{k+q}})=-2E_Fl^2.
$$

Аналогично,
$$
E_{\mathbf{k}}-E_{\mathbf{k-q}}-\hbar(\omega+i \nu)=2E_Fl(P_x-
\dfrac{l}{2}-\dfrac{z}{l}),
$$
$$
E_{\mathbf{k}}-E_{\mathbf{k+q}}+\hbar(\omega+i \nu)=-2E_Fl(P_x+
\dfrac{l}{2}-\dfrac{z}{l}).
$$

С помощью этих соотношений преобразуем интегралы (3.6) и (3.7).
Имеем:
$$
J_\omega=\dfrac{k_F^3}{8\pi^3N}\int \dfrac{f_{\mathbf{k}}P_\perp^2 d^3P}
{(P_x-z/l)^2-(l/2)^2},\qquad
J_\nu=\dfrac{k_F^3}{8\pi^3N}\int \dfrac{f_{\mathbf{k}}P_\perp^2 d^3P}
{P_x^2-(l/2)^2}.
$$

В этих выражениях
$$
z=\dfrac{\omega+i \nu}{k_Fv_F}=x+iy, \qquad
x=\dfrac{\omega}{k_Fv_F},\qquad y=\dfrac{\nu}{k_Fv_F}.
$$

Заметим, что для вырожденной плазмы выполняется соотношение
$k_F=3\pi^2N$, $f_{\mathbf{k}}=\Theta(E_F-E_{\mathbf{k}})=
\Theta(1-P^2)$. Поэтому интегралы (3.6) и (3.7) упрощаются:
$$
J_\omega=\dfrac{3}{8\pi}\int\dfrac{\Theta(1-P^2)P_\perp^2d^3P}
{(P_x-z/l)^2-(l/2)^2},\quad
J_\nu=\dfrac{3}{8\pi}\int\dfrac{\Theta(1-P^2)P_\perp^2d^3P}
{P_x^2-(l/2)^2}.
$$

Эти интегралы легко вычисляются:
$$
J_\omega=\dfrac{3}{8\pi}\int\limits_{P^2<1}\dfrac{P_\perp^2d^3P}
{(P_x-z/l)^2-(l/2)^2}=\dfrac{3}{16}\int\limits_{-1}^{1}
\dfrac{(1-t^2)^2dt}{(t-z/l)^2-(l/2)^2}=
$$
$$
=\dfrac{3}{16}\Bigg[
-\dfrac{10}{3}+\dfrac{4z^2}{l^2}+\dfrac{l^2}{2}+
\dfrac{1}{l}\Big[\Big(1-\dfrac{z^2}{l^2}\Big)^2+\dfrac{l^4}{16}-
\dfrac{l^2}{2}+\dfrac{3z^2}{2}\Big]\ln\dfrac{(1-l/2)^2-z^2/l^2}
{(1+z/l)^2-l^2/4}-
$$
$$
-\dfrac{zl}{2}\Big[1+\dfrac{4}{l^2}\Big(1-\dfrac{z^2}{l^2}\Big)\Big]
\ln\dfrac{(1-z/l)^2-l^2/4}{(1+z/l)^2-l^2/4}\Bigg],
$$
и
$$
J_\nu=\dfrac{3}{8\pi}\int \dfrac{P_\perp^2d^3P}{P_x^2-(l/2)^2}=
\dfrac{3}{8}\int\limits_{0}^{1}\dfrac{(1-t^2)^2dt}{t^2-(l/2)^2}=
$$
$$
=\dfrac{3}{8}\Bigg[-\dfrac{5}{3}+\dfrac{l^2}{4}+\dfrac{(l^2-4)^2}{16l}
\ln\dfrac{2-l}{2+l}\Bigg].
$$

Проверим выполнение одного из соотношений, называемого правилом $f$--сумм
(см., например, \cite{Dressel}, \cite{Pains} и \cite{Martin}) для поперечной
диэлектрической проницаемости (3.5). Это правило выражается формулой (4.200)
из монографии \cite{Pains}:
$$
\int\limits_{-\infty}^{\infty}\varepsilon_{tr}(\mathbf{q},\omega,\nu)\omega
d\omega=\pi \omega_p^2,
\eqno{(3.8)}
$$
где $\omega_p$ -- плазменная (ленгмюровская) частота,
$$
\omega_p=\sqrt{\dfrac{4\pi e^2 N}{m}}.
$$

Как показано в \cite{Pains}, для доказательства соотношения (4.1) достаточно
доказать выполнение предельного соотношения
$$
\varepsilon_{tr}(\mathbf{q},\omega,\nu)=1-\dfrac{\omega_p^2}{\omega^2}+
o\Big(\dfrac{1}{\omega^2}\Big), \qquad \omega\to \infty.
\eqno{(3.9)}
$$

Воспользуемся выражением (3.5) для поперечной электрической
проводимости. Из формулы (3.5) для проводимости и формул (3.6) и (3.7)
для интегралов $J_\omega$ и $J_\nu$ видно, что поперечная
проводимость удовлетворяет предельному соотношению (3.9).

Таким образом, правило $f$--сумм \cite{Pains} для поперечной диэлектрической
проницаемости квантовой столновительной плазмы выполняется.

Теперь представим формулу (3.4) для поперечной проводимости в
интегральной форме
$$
\dfrac{\sigma_{tr}}{\sigma_0}=\dfrac{iy}{x}-\dfrac{3y^2}{8x^2(1+iy/x)}
\int\limits_{0}^{1}\dfrac{(1-t^2)^2dt}{t^2-q^2/4}+$$$$+
\dfrac{3iy}{16(x+iy)}\int\limits_{-1}^{1}
\dfrac{(1-t^2)^2dt}{(t-z/q)^2-q^2/4}.
$$

Перепишем формулы (3.4) и (3.5) в безразмерных параметрах:
$$
\dfrac{\sigma_{tr}}{\sigma_0}=\dfrac{iy}{x}\Big(1+\dfrac{xJ_\omega+
iyJ_\nu}{x+iy}\Big)
$$
и
$$
\varepsilon_{tr}=1-\dfrac{x_p^2}{x^2}\Big(1+\dfrac{xJ_\omega+
iyJ_\nu}{x+iy}\Big).
$$

Здесь $x_p=\dfrac{\omega}{k_Fv_F}$ -- безразмерная плазменная
(ленгмюровская) частота.

Для сравнения представим формулу Линдхарда \cite{Lin} в наших обозначениях
следующим образом
$$
\dfrac{\sigma_{tr}^{Lin}}{\sigma_0}=\dfrac{iy}{x}(1+J_\omega).
\eqno{(3.10)}
$$

Из соотношений (3.4) и (3.10) видно, что мнимые части поперечных электрических
проводимостей из настоящей работы и из работы Линдхарда
при $y\to 0$ совпадают:
$$
\lim\limits_{y\to 0}\Im \sigma_{tr}=
\Im \sigma_{tr}^{Lin},\qquad \Im \sigma_{tr}^{Lin}=\dfrac{e^2N}{m\omega}
(1+J_\omega\Big|_{y=0}).
$$

Представим формулу для поперечной электрической проводимости в явном виде:
$$
\dfrac{\sigma_{tr}}{\sigma_0}=\dfrac{iy}{x}\Bigg\{1+\dfrac{3iy}{x}
\Big[-\dfrac{5}{3}+
\dfrac{q^2}{4}+\dfrac{(q^2-4)^2}{16q}\ln\dfrac{2-q}{2+q}\Big]+
\dfrac{3x}{16z}\bigg[
-\dfrac{10}{3}+\dfrac{4z^2}{q^2}+
$$
$$
+\dfrac{q^2}{2}+\dfrac{1}{q}\Big[\Big(1-\dfrac{z^2}{q^2}\Big)^2+\dfrac{q^4}{16}-
\dfrac{q^2}{2}+\dfrac{3z^2}{2}\Big]\ln\dfrac{(1-q/2)^2-z^2/q^2}
{(1+z/q)^2-q^2/4}-
$$
$$
-\dfrac{zq}{2}\Big[1+\dfrac{4}{q^2}\Big(1-\dfrac{z^2}{q^2}\Big)\Big]
\ln\dfrac{(1-z/q)^2-q^2/4}{(1+z/q)^2-q^2/4}
\bigg]\Bigg\}.
$$

\begin{center}
  \bf 4. Частные случаи электрической проводимости
\end{center}

Исследуем частные случаи электрической проводимости.
Возьмем формулу (3.4) и преобразуем ее к виду
$$
\dfrac{\sigma_{tr}}{\sigma_0}=\dfrac{i \nu}{\omega}\Bigg[1+
\dfrac{3i\nu}{8\pi(\omega+i \nu)q}
\int\dfrac{\Theta(1-P^2)-\Theta(1-(\mathbf{P-q})^2)}{P_x-q/2}P_\perp^2
d^3P+
$$
$$
+\dfrac{3\omega}{8\pi(\omega+i \nu)q}\int
\dfrac{\Theta(1-P^2)-\Theta(1-(\mathbf{P-q})^2)}
{P_x-z/q-q/2}P_\perp^2d^3P\Bigg].
\eqno{(4.1)}
$$

Подынтегральные выражения в этих интегралах содержат функцию
$$
\varphi(\mathbf{q})=\Theta(1-P^2)-\Theta(1-(\mathbf{P-q})^2).
$$

В линейном приближении мы получаем
$$
\varphi(\mathbf{q})=-2\delta(1-P^2)P_xq=-\delta(1-P)P_xq.
$$
Теперь выражение (4.1) упрощается
$$
\dfrac{\sigma_{tr}}{\sigma_0}=\dfrac{i \nu}{\omega}\Bigg[1-\dfrac{3i\nu}
{8\pi(\omega+i \nu)}
\int\dfrac{P_x\delta(1-P)}{P_x-q/2}P_\perp^2d^3P-
$$
$$
-\dfrac{3\omega}{8\pi(\omega+i \nu)}\int
\dfrac{P_x\delta(1-P)P_\perp^2d^3P}{P_x-z/q-q/2}\Bigg].
\eqno{(4.2)}
$$
Заметим, что
$$
\dfrac{3}{8\pi}\int \delta(1-P)(P^2-P_x^2)d^3P=1.
\eqno{(4.3)}
$$

Следовательно, выражение (4.2) может быть преобразовано к следующему виду
$$
\dfrac{\sigma_{tr}}{\sigma_0}=\dfrac{i \nu}{\omega}\Bigg[
\dfrac{\omega}{\omega+i \nu}-\dfrac{3i\nu q}
{16\pi(\omega+i \nu)}
\int\dfrac{\delta(1-P)(P^2-P_x^2)d^3P}{P_x-q/2}-
$$
$$
-\dfrac{3\omega}{8\pi(\omega+i \nu)}\int
\dfrac{P_x\delta(1-P)(P^2-P_x^2)d^3P}
{P_x-z/q-q/2}\Bigg].
\eqno{(4.4)}
$$

Теперь из формулы (4.4) видно, что при $q\to 0$
$$
\sigma_{tr}=\dfrac{i \nu}{\omega+i \nu}\sigma_0,
$$
и при $\omega=0$ мы получаем в точности статическую проводимость:
$\sigma_{tr}=\sigma_0$.

Покажем теперь, что при малых $q$ выражение (4.4) приводит к известному
выражению для проводимости вырожденной фермиевской плазмы.
В самом деле, заметим, что при малых $q$ первый интеграл из (4.4)
пропорционален $q^2$.Отбросим этот интеграл. В заменателе второго интеграла
пренебрегаем членом $q/2$, ибо $q/2 \ll |z|/q $.
В результате для малых $q$ получаем:
$$
\dfrac{\sigma_{tr}}{\sigma_0}=\dfrac{i \nu}{\omega + i
\nu}\Bigg[1-\dfrac{3}{8\pi}\int\dfrac{\delta(1-P)P_x}{P_x-z/q}
(P^2-P_x^2)d^3P\Bigg].
\eqno{(4.5)}
$$

Пользуясь снова равенством (4.3), на основании (4.5) приходим к
следующему выражению
$$
\dfrac{\sigma_{tr}}{\sigma_0}=-\dfrac{3y}{8\pi q}\int \dfrac{\delta(1-P)
(P^2-P_x^2)}{P_x-q/2}d^3P,
$$
которое приводит к известному выражению электрической
проводимости в вырожденной фермиевской плазме
$$
\sigma_{tr}^{\rm classic}=-\dfrac{3iy\sigma_0}{4q}
\int\limits_{-1}^{1}\dfrac{1-\mu^2}{\mu-z}d\mu=
\dfrac{3iy\sigma_0}{4q}\Big[\dfrac{2z}{q}+(z^2-q^2)\ln\dfrac{z-q}{z+q}\Big].
\eqno{(4.6)}
$$

На рис. 1 -- 8 представим сравнение модулей действительной и мнимой
частей трех электрических проводимостей. Одна из этих проводимостей
введенна в настоящей работе (это формула (3.4) и кривые "1"\, на рисунках),
вторая --- это проводимость Линдхарда (формула (3.10) и кривые "2"), третья
проводимость --- это классическая проводимость (формула (4.6) и
кривые "3").
Представим на рис. 9 и 10 зависимость действительной и мнимой частей
электрической проводимости, введенной в настоящей работе, от
безразмерной частоты колебаний $\omega$ при различных значениях
безразмерного волного числа $q$.

\begin{center}
\bf 5. Заключение
\end{center}

В настоящей работе выведена формула для электрической
проводимости в квантовой столкновительной плазме.
Для этой цели используется кинетическое уравнение с интегралом столкновений
в форме релаксационной модели в пространстве импульсов.
Выделяется и исследуется случай вырожденной фермиевской плазмы.
Проводится графическое сравнение проводимости из настоящей работы с
проводимостью по Линдхарду и с классической проводимостью.
\begin{figure}[h]
\begin{center}
\includegraphics[width=17.0cm, height=12cm]{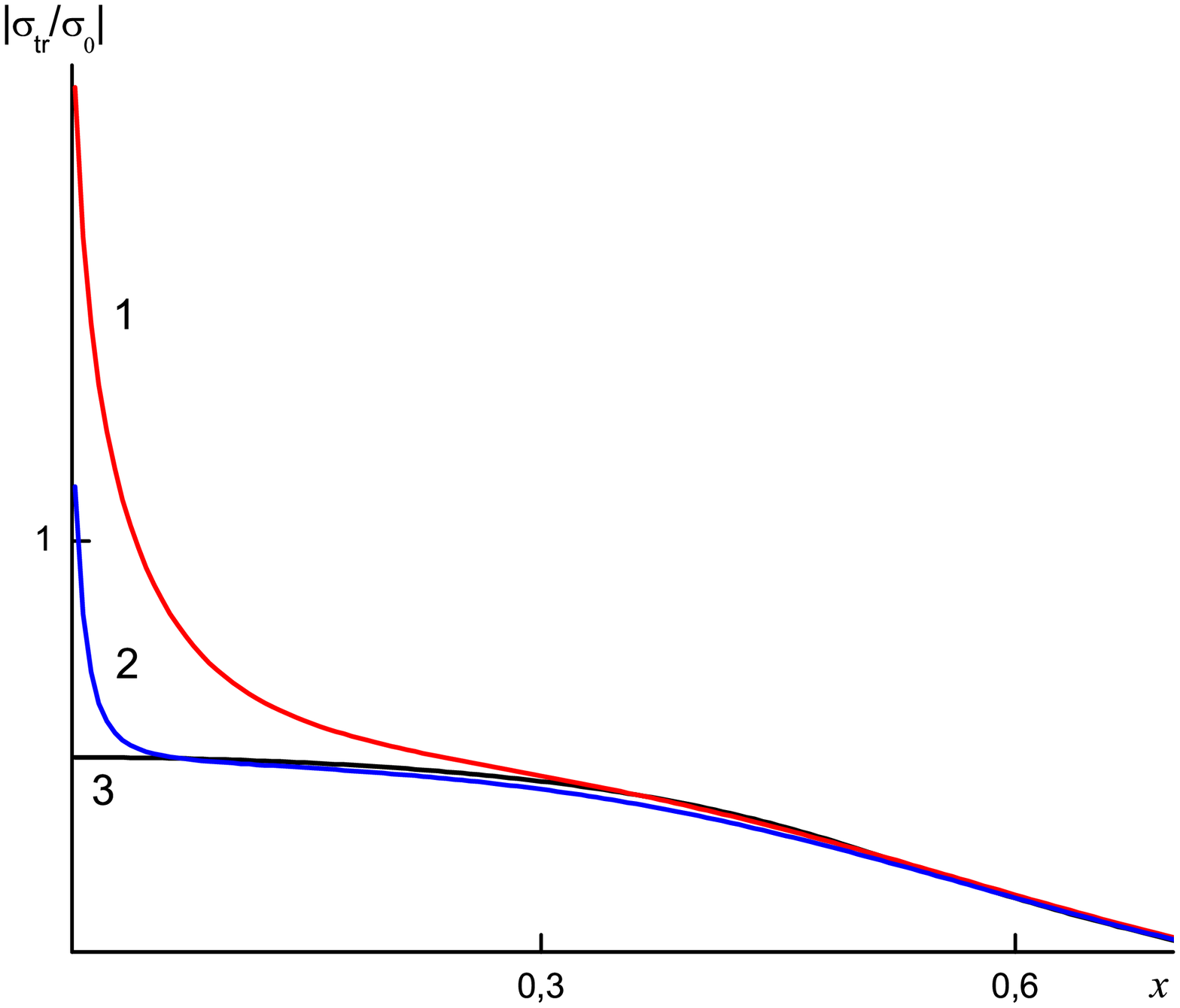}
\end{center}
\begin{center}
{{ Fig. 1. Dependence of $|\sigma_{tr}/\sigma_0|$ on quantity $x$; $y=0.1,
q=0.5$.}}
\end{center}
\end{figure}

\begin{figure}[h]
\begin{center}
\includegraphics[width=17.0cm, height=8cm]{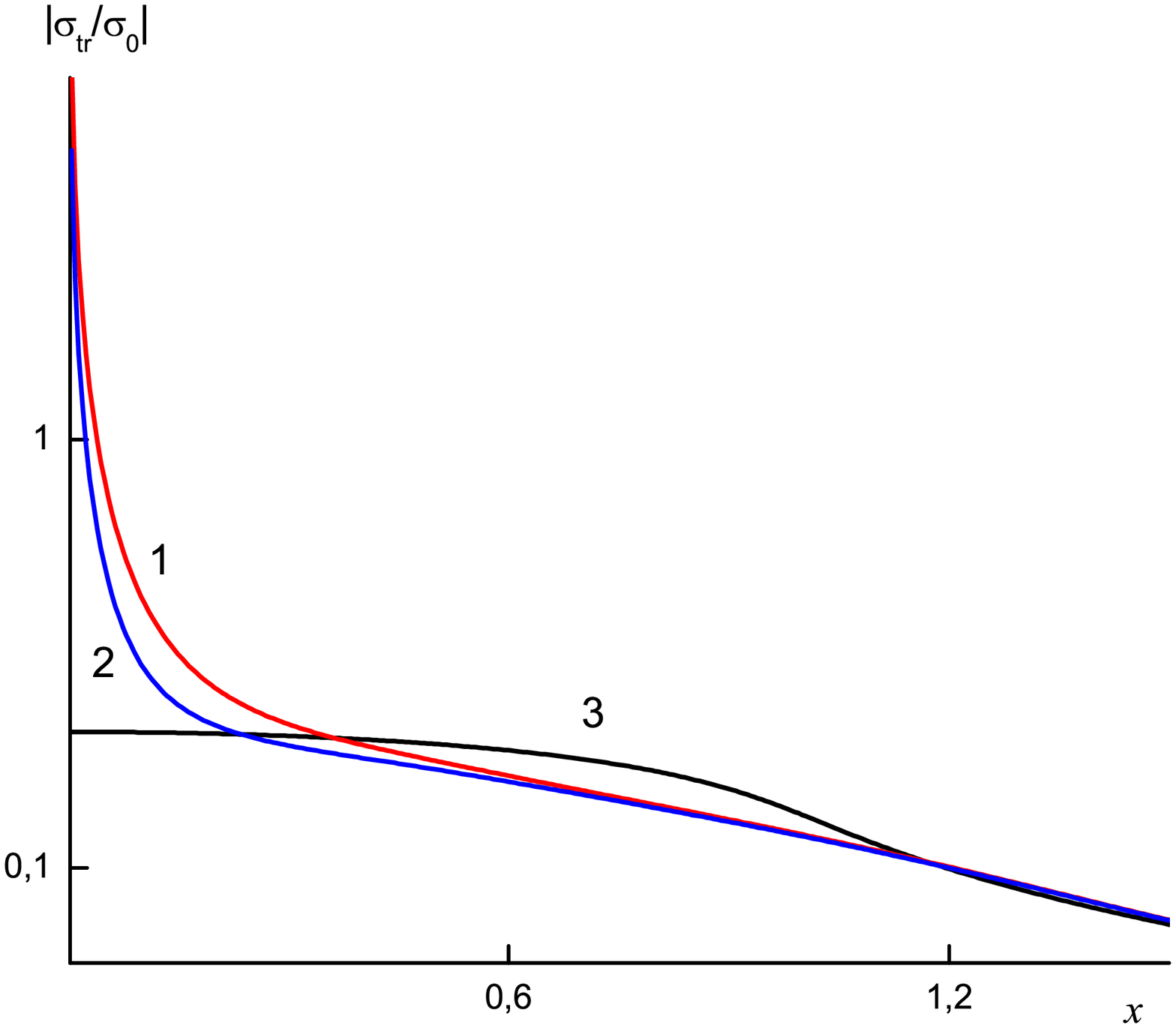}
\end{center}
\begin{center}
{{ Fig. 2. Dependence $|\sigma_{tr}/\sigma_0|$ of quantity $x$; $y=0.1, q=1$.}}
\end{center}
\end{figure}

\begin{figure}[h]
\begin{center}
\includegraphics[width=17.0cm, height=7cm]{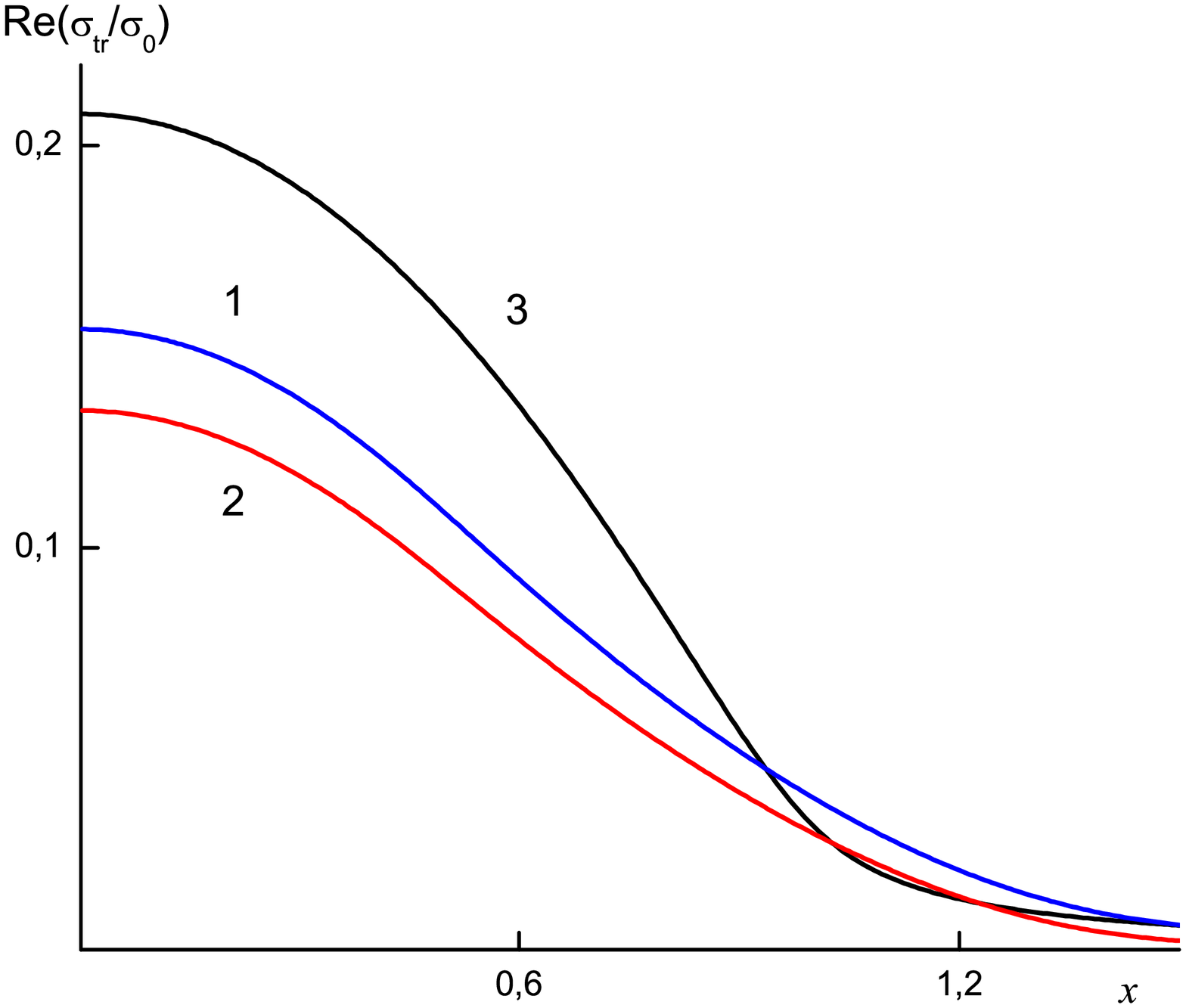}
\end{center}
\begin{center}
{{ Fig. 3. Dependence of $\Re(\sigma_{tr}/\sigma_0)$ on quantity $x$; $y=0.1,
q=1$.}}
\end{center}
\end{figure}

\begin{figure}[h]
\begin{center}
\includegraphics[width=17.0cm, height=9cm]{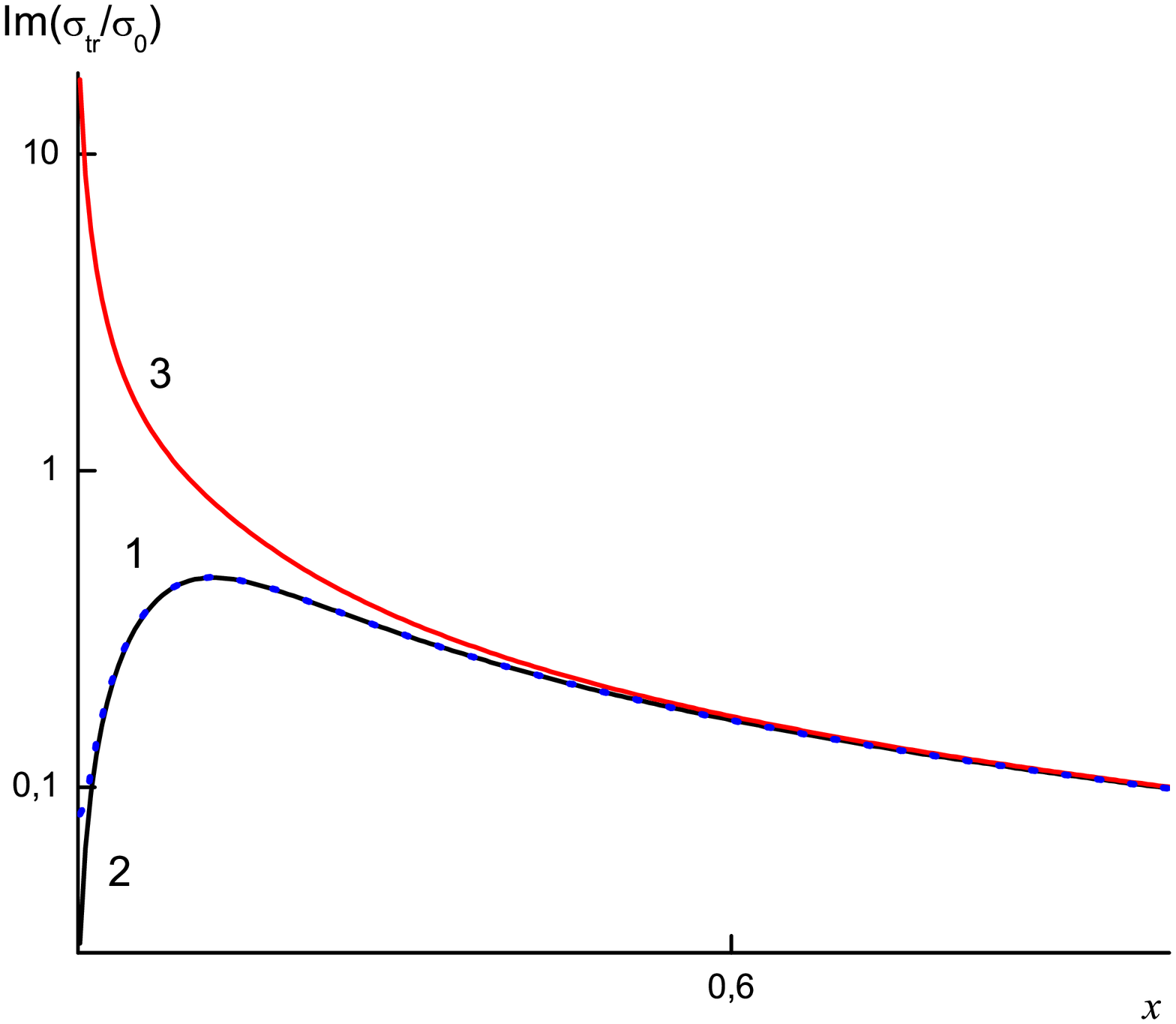}
\end{center}
\begin{center}
{{ Fig. 4. Dependence of $\Im(\sigma_{tr}/\sigma_0)$ on quantity $x$; $y=0.1, q=1$.}}
\end{center}
\end{figure}

\begin{figure}[h]
\begin{center}
\includegraphics[width=17.0cm, height=9cm]{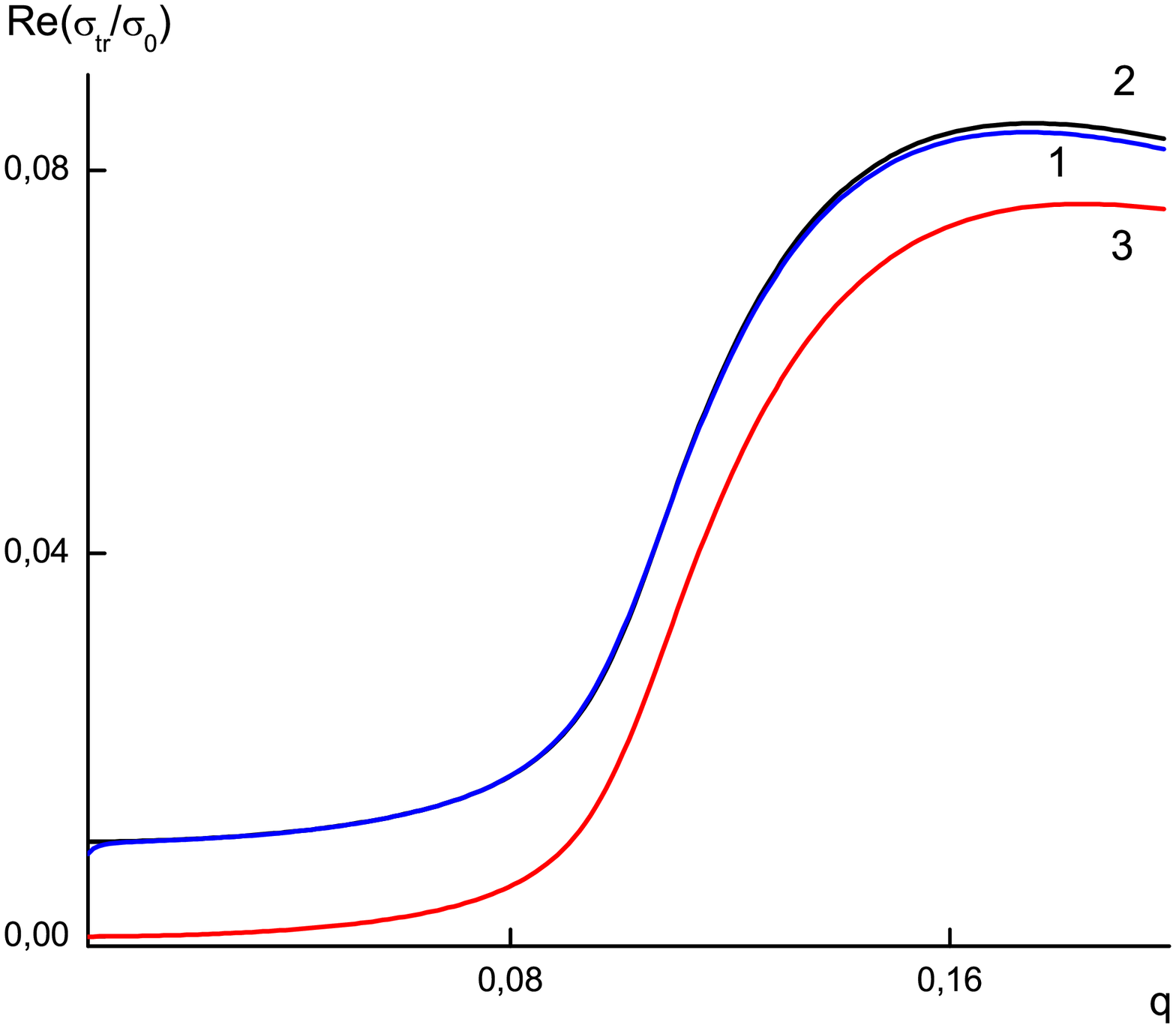}
\end{center}
\begin{center}
{{ Fig. 5. Dependence of $\Re(\sigma_{tr}/\sigma_0)$ on quantity $q$;
$x=0.1, y=0.01, 0\leqslant q \leqslant 0.2$.}}
\end{center}
\end{figure}

\begin{figure}[h]
\begin{center}
\includegraphics[width=17.0cm, height=9cm]{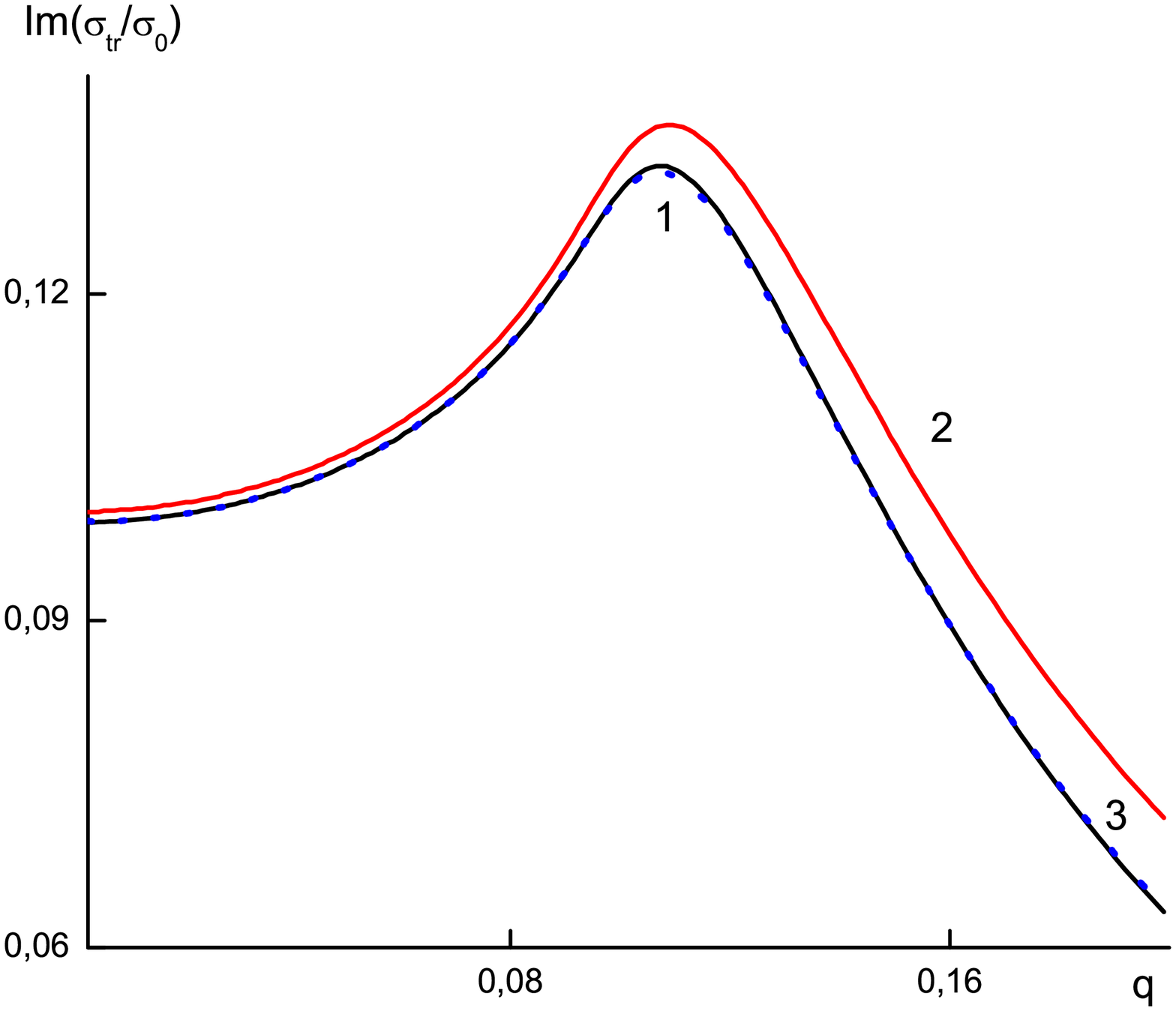}
\end{center}
\begin{center}
{{ Fig. 6. Dependence of $\Im(\sigma_{tr}/\sigma_0)$ on quantity $q$;
$x=0.1, y=0.01, 0\leqslant q \leqslant 0.2$.}}
\end{center}
\end{figure}

\begin{figure}[h]
\begin{center}
\includegraphics[width=17.0cm, height=9cm]{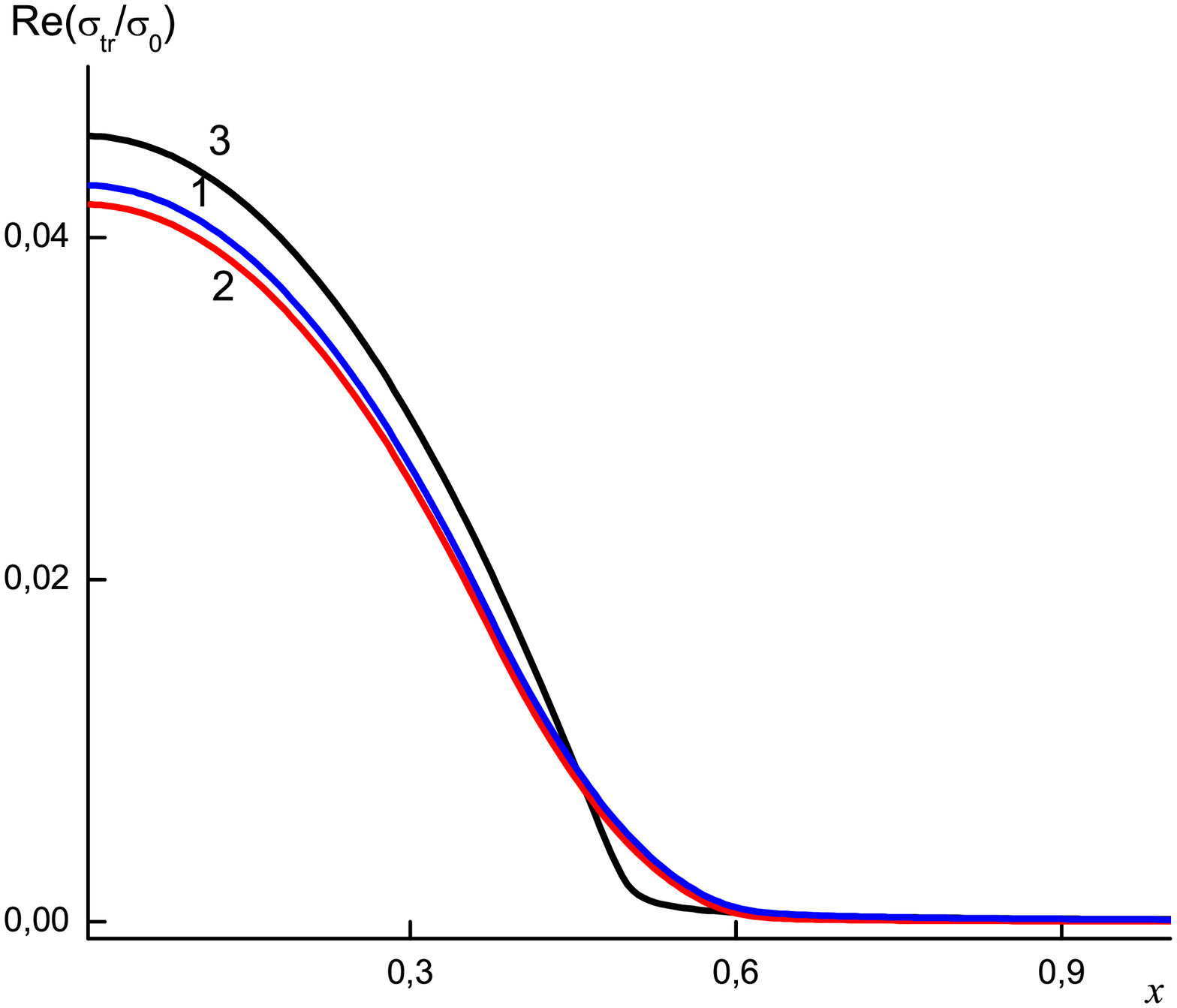}
\end{center}
\begin{center}
{{ Fig. 7. Dependence of $\Re(\sigma_{tr}/\sigma_0)$ on quantity $x$;
$q=0.5, y=0.01$.}}
\end{center}
\end{figure}

\begin{figure}[h]
\begin{center}
\includegraphics[width=17.0cm, height=9cm]{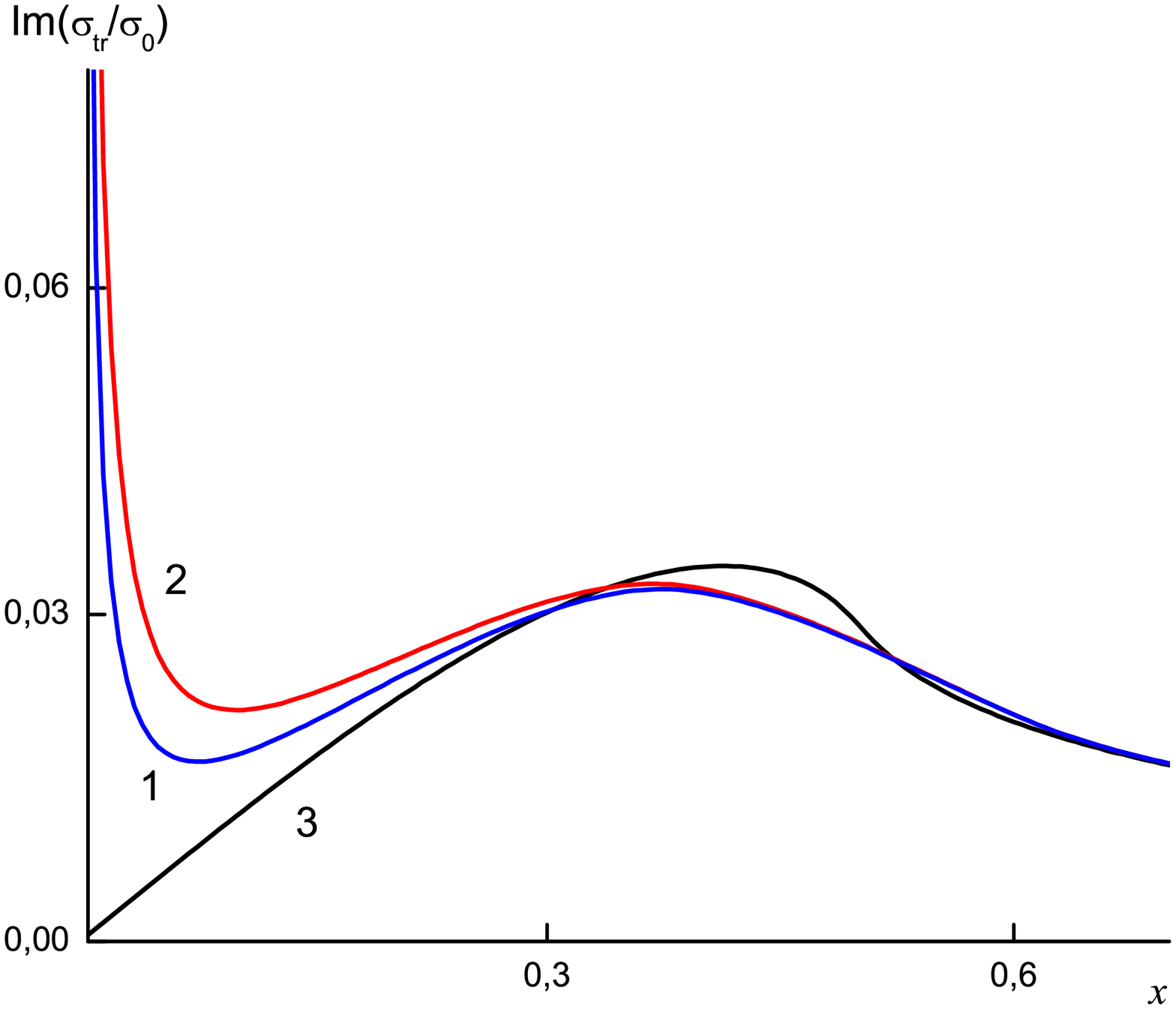}
\end{center}
\begin{center}
{{ Fig. 8. Dependence of $\Im(\sigma_{tr}/\sigma_0)$ on quantity $x$;
$q=0.5, y=0.01$.}}
\end{center}
\end{figure}

\begin{figure}[h]
\begin{center}
\includegraphics[width=17.0cm, height=8cm]{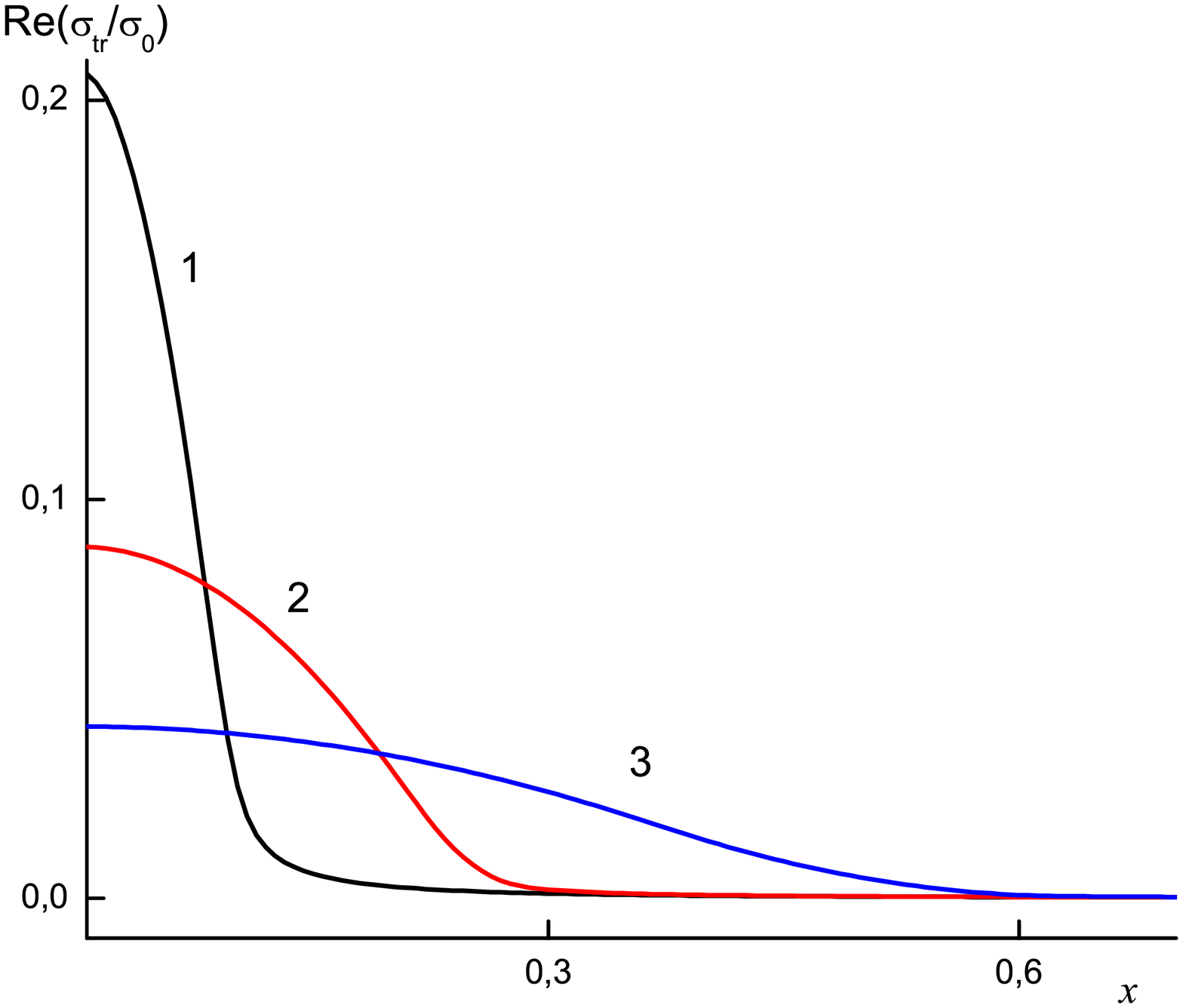}
\end{center}
\begin{center}
{{ Fig. 9. Dependence of $\Re(\sigma_{tr}/\sigma_0)$ on quantity $x$;
$y=0.01$, сurves of $1,2,3$ correspond to values of parameter $q$: $q=0.1, 0.25,
0.5$.}}
\end{center}
\end{figure}

\begin{figure}[h]
\begin{center}
\includegraphics[width=17.0cm, height=8cm]{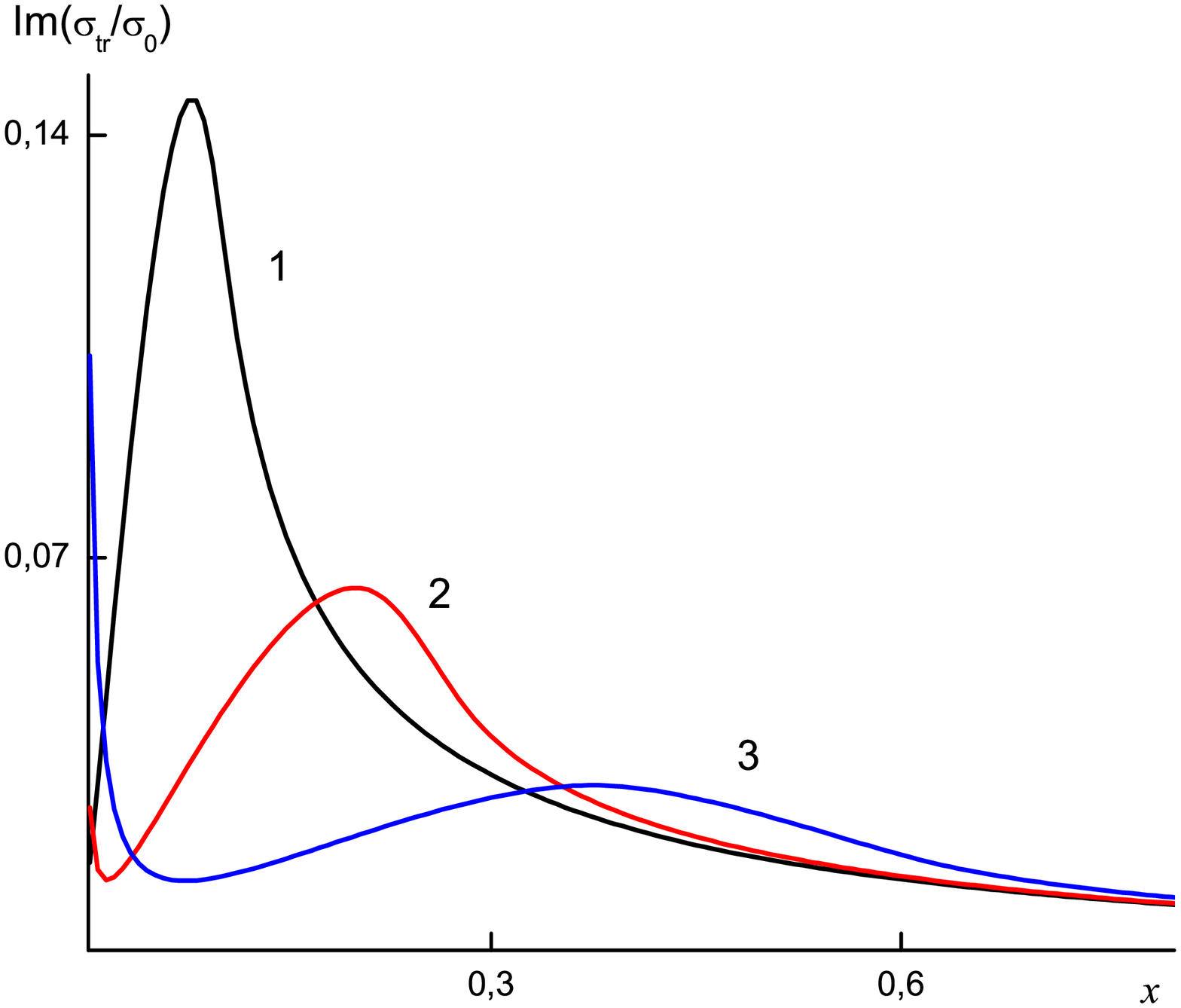}
\end{center}
\begin{center}
{{ Fig. 10. Dependence of $\Im(\sigma_{tr}/\sigma_0)$ on quantity $x$;
$y=0.01$, curves $1,2,3$ correspond to values of parameter $q$: $q=0.1, 0.25,
0.5$.}}
\end{center}
\end{figure}

\end{document}